\documentclass[11pt,twoside]{article}

\usepackage{asp2006}
\usepackage{epsf}
\usepackage{graphicx}
\usepackage{lscape}

\begin{document}
\title{The Contribution of Radio Selected Star Forming Galaxies to the IR 
  Energy Density Budget}   
\author{N. Seymour\altaffilmark{1}, T. Dwelly\altaffilmark{2}, D. Moss\altaffilmark{2}, I. M$^{\rm c}$Hardy\altaffilmark{2}, A. Zoghbi\altaffilmark{2,3}, G. Rieke\altaffilmark{4}, M. Page\altaffilmark{5}, A. Hopkins\altaffilmark{6}}
\altaffiltext{1}{{\it Spitzer} Science Center}
\altaffiltext{2}{University of Southampton}
\altaffiltext{3}{Institute of Astronomy, Cambridge}
\altaffiltext{4}{Steward Observatory}
\altaffiltext{5}{MSSL}
\altaffiltext{6}{University of Sydney}

\begin{abstract} 
We have used several different methods (radio morphology, 
radio spectral index, mid-IR to radio and near-IR to radio flux density 
ratios) to discriminate between AGN and 
SFGs in faint, sub-mJy radio surveys. We find that the latter two 
methods are the most powerful with current multi-wavelength data, 
but that future radio surveys with eMERLIN, LOFAR etc. (and ultimately the 
SKA) will greatly increase
the power of the morphology and spectral index methods. As an example of the 
science possible we derive the IR luminosity density from the radio-selected 
SFGs using the radio/IR
luminosity correlation. We also examine the contribution by luminosity to the
total IR luminosity density and find evidence that fraction of LIRGs 
remains constant or decreases above $z=1$ while the relative fraction of 
ULIRGs continues to increase up to $z=2.5$.

\end{abstract}


\section{Discriminating Between AGN and SFGs}

Deep extragalactic radio surveys have largely remained under-exploited due to 
the difficulty in discriminating between AGN and SFGs at the faintest flux 
densities. This situation 
is now ameliorated by deep multi-wavelength surveys where we can begin
to separate these populations for the first time. Obtaining large samples of 
radio-selected SFGs and AGN is very important for our understanding of
galaxy evolution. Whereas most tracers of star formation may be obscured by 
gas or dust, radio emission is a direct tracer of SFR and the SKA, for example,
will be able to detect $L_*$ galaxies, like the Milky Way, out to the era of 
reionization.
Radio-selected AGN are also important as they are an unobscured marker of 
accretion onto a super-massive black hole. Either radio-loud AGN have exactly 
the same properties as radio-quiet AGN (bar the radio jets, Urry \& Padovani 
1992) and hence can be an unbiased measure of AGN properties and activity, 
or the radio jets 
play an important role in the evolution of the AGN (Saunders et al. 1999) 
and well selected samples of radio-loud and radio-quiet AGN 
are needed to understand the role of radio jets in AGN evolution.

\begin{figure}[!ht]
\centering	
\includegraphics[width=3.2in,angle=270]{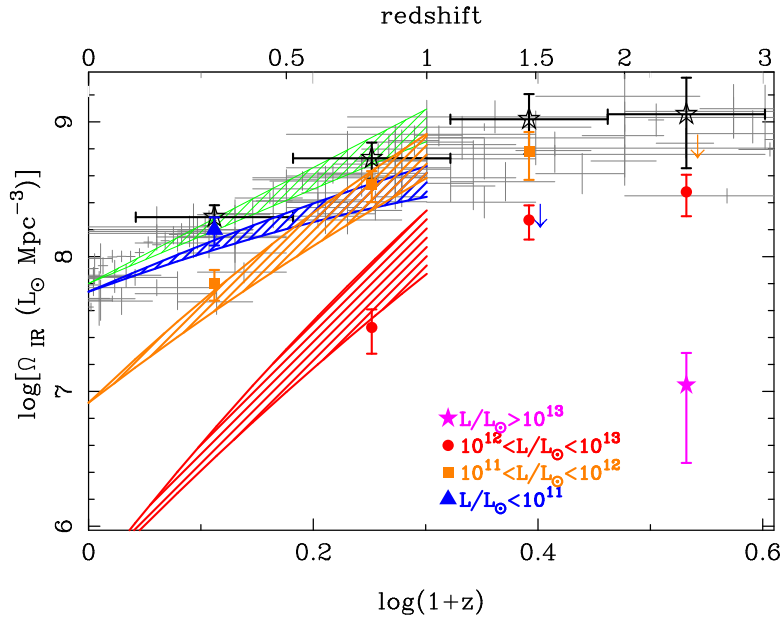}
\caption{We plot the total comoving IR luminosity density as a function of 
redshift. 
The shaded regions are from Le Floc'h et al. (2005) and the points are from
this work adapted from the radio-selected SFGs presented in Seymour et al. 
(2008). Both samples have 
their total IR density (green region and open stars respectively)
subdivided by luminosity into sub-LIRGs, LIRGs, ULIRGs and HyLIRGs as 
indicated by the legend.}
\label{fig1}
\end{figure}

\section{Total IR Luminosity Density}

We have previously used the radio-selected SFG population to derive the 
comoving SFR density of the Universe up to $z\sim2.5$. This result agrees well
with those measurements at other wavelengths (Seymour et al. 2008). Using the
radio/total-IR luminosity relation (Bell 2003) we can convert the radio 
luminosity to IR luminosity and calculate the IR luminosity density if the 
Universe as a function of redshift. We compare our result to that of 
Le Floc'h et al. (2005) in Figure~\ref{fig1}, but extend our results to higher
redshifts. We are also able to determine the contribution by luminosity bin
at each epoch and find that more luminous sources continue to dominate more at 
higher redshifts.


\acknowledgements 
NS would like to thank Taddy and the rest of the organizers for a wonderful 
conference and Emeric Le Floc'h for useful discussions.


\end{document}